\documentclass[a4paper,fleqn]{cas-dc}
\usepackage{cite}
\usepackage{amsmath,amssymb,amsfonts}
\usepackage{bm}
\usepackage{algorithm}
\usepackage{algorithmic}
\usepackage{graphicx}
\usepackage{textcomp}
\usepackage{multirow}
\usepackage{booktabs}
\usepackage[utf8]{inputenc}

\usepackage{etoolbox}
\usepackage{bm}

\usepackage{amsthm}%liwen
\usepackage{algorithm}%liwen
\usepackage{algorithmic}%liwen
\usepackage{subfigure}%liwen
\usepackage{xspace}
\usepackage{url}

\newcommand{\MyMapTemplatePrefixc}[4]{\expandafter#1\csname#3#4\endcsname{#2{#4}}} % it remembles a template: \#3#4 --> #2{#4}
\forcsvlist{\MyMapTemplatePrefixc {\def} {\mathcal}{c}} {A,B,C,D,E,F,G,H,I,J,K,L,M,N,O,P,Q,R,S,T,U,V,W,X,Y,Z}  % e.g., \cA --> \mathcal{A}

\newcommand{\MyMapTemplatePrefixtb}[5]{\expandafter#1\csname#4#5\endcsname{#2{#3{#5}}}} % it remembles a template: \#3#4 --> #2{#4}
\forcsvlist{\MyMapTemplatePrefixtb {\def} {\tilde}{\mathbf}{t}} {A,B,C,D,E,F,G,H,I,J,K,L,M,N,O,P,Q,R,S,T,U,V,W,X,Y,Z}  % e.g., \tA --> \tilde{A}
\forcsvlist{\MyMapTemplatePrefixtb {\def} {\tilde}{\mathbf}{t}} {0,1,a,b,c,d,e,f,g,h,i,j,k,l,m,n,o,p,q,r,s,u,v,w,x,y,z}  % e.g., \ta --> \tilde{a}

\newcommand{\MyMapTemplateNoPrefix}[3]{\expandafter#1\csname#3\endcsname{#2{#3}}}
\forcsvlist{\MyMapTemplateNoPrefix {\def} {\mathbf} } {0,1,a,b,d,e, f, g, h, i, j, k, l, m, n, o, p, q, r, u, w, x, y, z} % e.g., \a --> \mathbf{a}, except for c and v for special characters
\forcsvlist{\MyMapTemplateNoPrefix {\def} {\mathbf} } {A,B,C,D,E,F,G,H,I,J,K,L,M,N,O,P,Q,R,S,T,U,V,W,X,Y,Z}  % e.g., \A --> \mathcal{A}

\def\etal{\emph{et al.}\@\xspace}
\def\ie{\emph{i.e.}\@\xspace}
\def\eg{\emph{e.g.}\@\xspace}

\usepackage[numbers,sort]{natbib}
\def\tsc#1{\csdef{#1}{\textsc{\lowercase{#1}}\xspace}}
\tsc{WGM}
\tsc{QE}
\begin{document}
\let\WriteBookmarks\relax
\def\floatpagepagefraction{1}
\def\textpagefraction{.001}

\shorttitle{A Federated Learning Approach with Noise-Resilient Training for MS Lesion Segmentation}

\shortauthors{L. Bai and D. Wang \etal}  

\title[mode = title]{Improving Multiple Sclerosis Lesion Segmentation Across Clinical Sites: A Federated Learning Approach with Noise-Resilient Training}

\author[1,2]{Lei Bai}
\fnmark[1]
\cormark[1]
% \ead{bailei@pjlab.org.cn}
\ead{baisanshi@gmail.com}
% % URL of the first author
% \ead[url]{http://leibai.site/}
% \cortext[1]{Equal contribution}
% % Credit authorship

\author[1,3]{Dongang Wang}[orcid=0000-0001-5805-0244]
\ead{dongang.wang@sydney.edu.au}
\cormark[1]
% \fntext[2]{This work was done when Lei Bai was working in the University of Sydney, Australia}

\author[1,3,4]{Michael Barnett}[orcid=0000-0002-2156-8864]
\author[1]{Mariano Cabezas}[orcid=0000-0002-4417-1704]
\author[1,5]{Weidong Cai}[orcid=0000-0003-3706-8896]
\author[1,6,7]{Fernando Calamante}[orcid=0000-0002-7550-3142]
\author[1,3]{Kain Kyle}
\author[1,5]{Dongnan Liu}[orcid=0000-0001-8102-3949]
\author[3]{Linda Ly}
\author[3]{Aria Nguyen}[orcid=0009-0003-0106-2721]
\fnmark[2]
\author[3]{Chun-Chien Shieh}[orcid=0000-0001-8245-8936]
\author[6,8]{Ryan Sullivan}[orcid=0000-0001-5554-7378]
\author[3]{Hengrui Wang}
% \ead{hengrui.wang@snac.com.au}
\author[1,3]{Geng Zhan}
\author[2]{Wanli Ouyang}[orcid=0000-0002-9163-2761]
\fnmark[1]
\cormark[2]
\author[1,3]{Chenyu Wang}[orcid=0000-0001-7135-7662]
\cormark[2]
\fnmark[3]
\ead{chenyu.wang@sydney.edu.au}

\affiliation[1]{organization={Brain and Mind Centre},
            addressline={The University of Sydney}, 
            city={NSW},
            citysep={},
            postcode={2050}, 
            country={Australia}}
\affiliation[2]{organization={School of Electrical and Information Engineering},
            addressline={The University of Sydney}, 
            city={NSW},
            citysep={},
            postcode={2006},
            country={Australia}}
\affiliation[3]{organization={Sydney Neuroimaging Analysis Centre},
            addressline={94 Mallett Street}, 
            city={NSW},
            citysep={},
            postcode={2050}, 
            country={Australia}}
\affiliation[4]{organization={Royal Prince Alfred Hospital},
            city={NSW},
            % citysep={},
            postcode={2050}, 
            country={Australia}}
\affiliation[5]{organization={School of Computer Science},
            addressline={The University of Sydney}, 
            city={NSW},
            citysep={},
            postcode={2006}, 
            country={Australia}}
\affiliation[6]{organization={School of Biomedical Engineering},
            addressline={The University of Sydney}, 
            city={NSW},
            citysep={},
            postcode={2006}, 
            country={Australia}}
\affiliation[7]{organization={Sydney Imaging},
            addressline={The University of Sydney}, 
            city={NSW},
            citysep={},
            postcode={2006}, 
            country={Australia}}
\affiliation[8]{organization={Australian Imaging Service},
            % addressline={The University of Sydney}, 
            city={NSW},
            citysep={},
            postcode={2006}, 
            country={Australia}}
            
% cormarkers
\cortext[1]{Equal contributed first authors}
\cortext[2]{Equal contributed senior authors}

\fntext[1]{The authors are now at Shanghai AI Laboratory, Shanghai, China.}
\fntext[2]{The author is now at School of Physics, The University of Sydney, Australia}
\fntext[3]{Corresponding author}
% \nonumnote{}

% \thanks{The paper is submitted for review on 30-July-2022. 
% ``This work was supported in part by the Australian Department of Health under Grant GA89125.'' }
% \thanks{Lei Bai is with School of Electrical and Information Engineering, the University of Sydney, NSW, 2006, Australia, and also with Brain and Mind
% Centre, The University of Sydney, Sydney 2050, Australia (email: baisanshi@gmail.com).}
% \thanks{Dongang Wang, Kain Kyle, Linda Ly, Chun-Chien Shieh, Aria Nguyen, Michael Barnett, and Chenyu Wang are with Sydney Neuroimaging Analysis Centre (SNAC), and Brain and Mind Centre, the University of Sydney, NSW, 2006, Australia.}
% \thanks{Dongnan Liu and Weidong Cai are with the School of Computer Science, The
% University of Sydney, NSW 2008, Australia, and also with Brain and Mind
% Centre, The University of Sydney, Sydney 2050, Australia.}
% \thanks{Geng Zhan and Mariano Cabezas are with Brain and Mind Centre, The University of Sydney, Sydney 2050, Australia.}
% \thanks{Fernando Calamante, is with the School of Biomedical Engineering, The
% University of Sydney, Sydney 2050, Australia, and also with Sydney
% Imaging, The University of Sydney, Sydney 2050, Australia}
% \thanks{Wanli Ouyang is with School of Electrical and Information Engineering, the University of Sydney, NSW, 2006, Australia (email: wanli.ouyang@sydney.edu.au).}
% }

\begin{abstract} 
Accurately measuring the evolution of Multiple Sclerosis (MS) with magnetic resonance imaging (MRI) critically informs understanding of disease progression and helps to direct therapeutic strategy. Deep learning models have shown promise for automatically segmenting MS lesions, but the scarcity of accurately annotated data hinders progress in this area. Obtaining sufficient data from a single clinical site is challenging and does not address the heterogeneous need for model robustness. Conversely, the collection of data from multiple sites introduces data privacy concerns and potential label noise due to varying annotation standards.
To address this dilemma, we explore the use of the federated learning framework while considering label noise. Our approach enables collaboration among multiple clinical sites without compromising data privacy under a federated learning paradigm that incorporates a noise-robust training strategy based on label correction.
Specifically, we introduce a Decoupled Hard Label Correction (DHLC) strategy that considers the imbalanced distribution and fuzzy boundaries of MS lesions, enabling the correction of false annotations based on prediction confidence. We also introduce a Centrally Enhanced Label Correction (CELC) strategy, which leverages the aggregated central model as a correction teacher for all sites, enhancing the reliability of the correction process.
Extensive experiments conducted on two multi-site datasets demonstrate the effectiveness and robustness of our proposed methods, indicating their potential for clinical applications in multi-site collaborations.

%Keep the abstract to 250 words or less.
\end{abstract}

% \begin{highlights}
% \item To the best of our knowledge, this work is the first to address the segmentation problem with noisy labels in a federated learning paradigm. 
% \item We propose a Decoupled Hard Label Correction (DHLC) strategy that automatically corrects imperfect annotations according to predictions of local models, modulated by the characteristics of the MS lesions.
% \item We propose a Centrally Enhanced Label Correction (CELC) strategy to train MS lesion segmentation models with noisy labels in the federated learning paradigm, which improves the proposed DHLC strategy by introducing the contribution of the central model. 
% \item We conducted extensive experiments on real clinical datasets with various types of artificial label noise and annotator-introduced label variations to demonstrate the effectiveness and robustness of our method. Trained with noisy labels in a federated learning paradigm, our method achieves comparable performance with a centralized model trained using the same dataset with clean labels.  
% \end{highlights}

% Keywords
% Each keyword is separated by \sep

\maketitle

\section{Introduction}
\label{sec:introduction}
Multiple sclerosis (MS) is a chronic neuroinflammatory disease that results in cumulative focal and diffuse damage to the brain and spinal cord. Magnetic Resonance Imaging (MRI) based volumetric quantitation of MS brain lesions, \ie T2 hyper-intense or T1 hypo-intense areas, and their change over time, informs both disease course and response to therapy~\cite{ma2022multiple}.

In current clinical trial workflows, MS lesions are manually segmented from images acquired over several time points by trained neuroimaging analysts to provide measures of therapeutic efficacy. 
However, annotating lesions on the several hundred images acquired in each scan is extremely time-consuming and labor-intensive, considering that the burden of MS lesions is highly heterogeneous, ranging from zero to hundreds in lesion number, and from less than 1ml to more than 15ml of brain tissue volume affected. Lesions can also be widely distributed throughout the brain, although they preferentially affect the periventricular, juxtacortical, and infratentorial regions.
Lack of analysis expertise, time constraints, and logistical problems preclude routine quantitative measurement of MS lesions in clinical practice, which is reliant upon qualitative interpretation of images by radiologists and clinicians.

To facilitate the annotation process to improve the current clinical workflow, both statistical methods and deep learning methods have been explored to automatically segment lesions, among which U-Net-based deep neural networks~\cite{ronneberger2015u,liu2023multiple} are most widely used in research settings.

The lack of large, high-quality multi-center datasets and corresponding ground truth annotations is a critical limitation for training deep neural networks in medical imaging tasks, including MS lesion segmentation. 
Ethical, governance, privacy, and security constraints are difficult to navigate and preclude the participation of many centers in conventional centralized learning frameworks, which require the aggregation of raw imaging into a single large-scale dataset. Federated Learning (FL), an emerging training framework that does not expose raw data from contributing sites, overcomes many of these obstacles. In FL, only the model weights, trained locally at each participating site, are shared with the central server for aggregation in an iterative training process that captures the distilled information of the site datasets without compromising raw data or private health information~\cite{mcmahan2017communication}.

\begin{figure}[!t]
\centerline{\includegraphics[width=0.9\columnwidth]{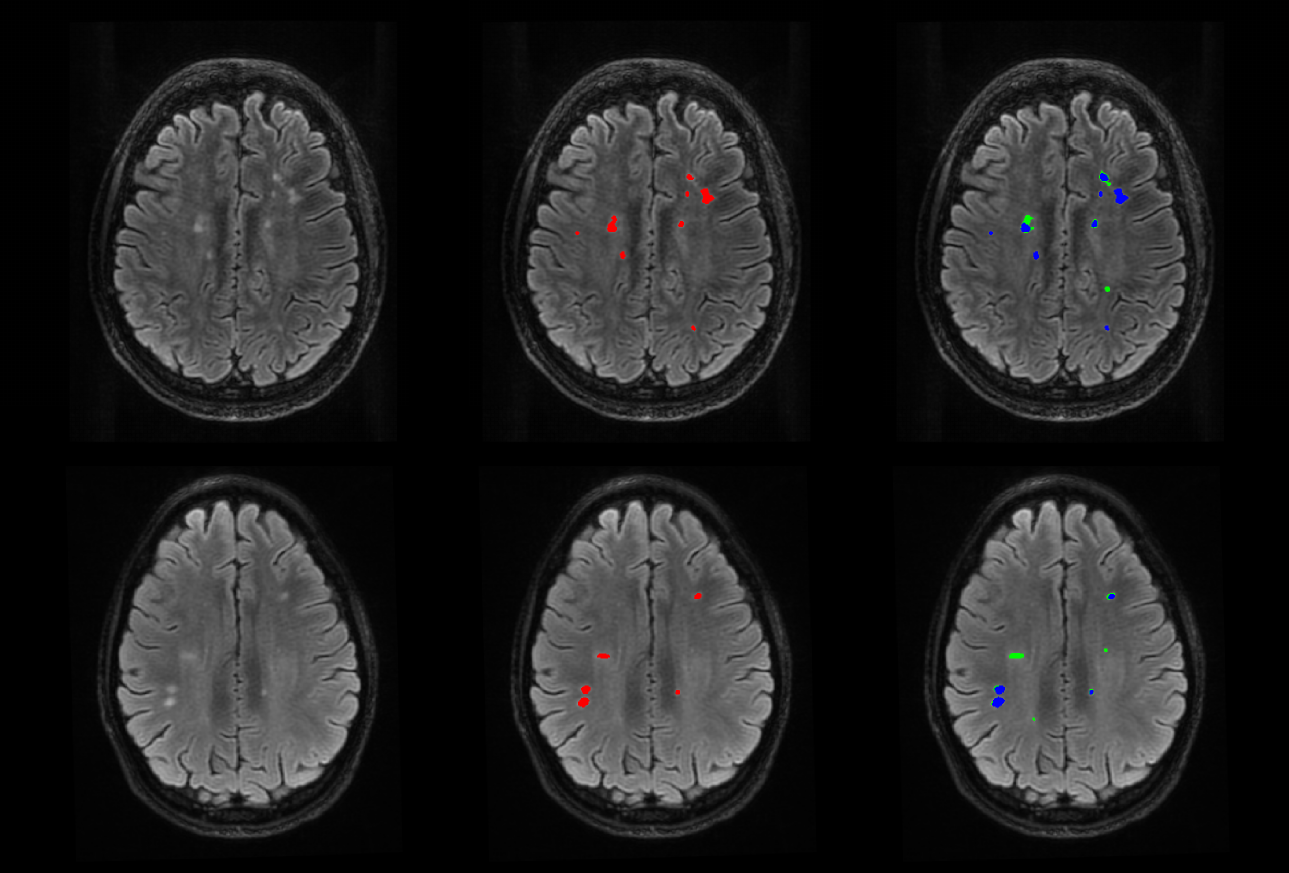}}
\caption{An example of noisy labels from different data centers. Each row represents one single case, and the three columns represent the original slices, consensus labels, and noisy labels respectively. The annotations in red are the consensus of several expert image analysts and the annotations in blue and green are from different annotators and are intended to illustrate the underestimation or overestimation of the lesions correspondingly. Better viewed in color.}
\label{fig:example}
\end{figure}

While FL is an effective training framework using scattered multi-center datasets and clean labels in both natural and medical image research~\cite{flores2021federated,ng2021federated}, the impact of inaccurate labels has not been explored for image segmentation tasks in a federated learning environment.
A multiplicity of image acquisition protocols and annotation schemes across participating centers generates significant labeling variability that is inevitably exacerbated by significant heterogeneity of MS lesion morphology and appearance~\cite{garcia2013review}. As shown in Figure~\ref{fig:example}, lesion segmentations performed on the same MRI images by two trained annotators differ from one another and from a consensus generated by the maximum voting of several expert image analysts, which is generally used to delineate ``golden'' labels for training and evaluation. Such label noise includes mostly the divergent boundaries of lesions and missed lesions according to our experience.

%Unlike the centralised training process, where all the data from different centers can be annotated by the same analysis centers and/or experts and the consensus annotations can be used as the ``golden'' labels for training and evaluation, the data from different sites may be annotated using local standards but used for training at the same time under the federated learning, which introduces inter-site and inter-expert variability, and the performance of the FL-based neural network model can be influenced accordingly~\cite{yang2022robust}.
%To the best of our knowledge, such an effect has not been studied for segmentation tasks, especially the MS lesion segmentation task.

In this work, we conduct the first exploration of MS lesion segmentation with noisy labels under a federated learning paradigm to achieve a more practical, scalable, and accurate automatic lesion annotation system. In contrast with existing works that employ centralized MS lesion segmentation training with consistent and clean labels~\cite{brosch2016deep,hou2019cross}, we consider a scenario in which labels from individual sites participating in a federated training framework are less controlled for annotation quality and propose a Decoupled Hard Label Correction (DHLC) strategy to cope with the label noise issue that takes into consideration the characteristics of MS lesions, \ie, the extreme imbalance number of voxels between the lesion and normal brain features, and the relatively fuzzier decision boundaries compared with other segmentation tasks (\eg, natural image segmentation and brain segmentation in the MRI images).

Moreover, we introduce a Centrally Enhanced Label Correction (CELC) strategy, in which the aggregation center in the federated training framework helps each participating site maintain a more accurate central model that can be utilized to decrease overfitting of each site model to their local label noise during the site updating process. 
Here, we experimentally demonstrate that noisy annotations severely degrade the performance of the deep models under a federated paradigm; and show that our proposed strategies (\ie, DHLC and CELC) can significantly improve the model's robustness against label noise. In the two multi-site datasets we explored in this study (\ie, the public MSSEG-2016 dataset and an in-house dataset SNAC-MS), our method consistently augments segmentation performance of the U-Net model trained with severe label noise.
%a novel method is proposed for MS lesion segmentation under the federated learning setting to deal with label noise. 

The contributions of our study are four-fold. 

\begin{enumerate}
\item To the best of our knowledge, this work is the first to address the segmentation problem with noisy labels in a federated learning paradigm. 
\item We propose a Decoupled Hard Label Correction (DHLC) strategy that automatically corrects imperfect annotations according to predictions of local models, modulated by the characteristics of the MS lesions.
\item We propose a Centrally Enhanced Label Correction (CELC) strategy to train MS lesion segmentation models with noisy labels in the federated learning paradigm, which improves the proposed DHLC strategy by introducing the contribution of the central model. 
\item We conducted extensive experiments on real clinical datasets with various types of artificial label noise and annotator-introduced label variations to demonstrate the effectiveness and robustness of our method. Trained with noisy labels in a federated learning paradigm, our method achieves comparable performance with a centralized model trained using the same dataset with clean labels. 
\end{enumerate}

\section{Related Works}\label{sec:relatework}
% In this section, literature relevant to our work is reviewed, including lesion segmentation and federated learning research, and corresponding methods related to noisy labels.

\subsection{Deep Learning based Lesion Segmentation}

MS lesion segmentation is based primarily upon altered signal intensity relative to normal white (and gray) matter on MRI sequences such as T1-weighted (T1-w), T2-weighted (T2-w), and T2 fluid-attenuated inversion recovery (T2-FLAIR) images. Deep learning based segmentation methods have shown promising performance. 

The earliest application of deep learning, described in~\cite{vaidya2015longitudinal} and~\cite{ghafoorian2017deep}, treated the segmentation problem as a voxel-wise classification problem, in which 3D convolutional neural networks with multiple layers and channels were used to predict the labels for each voxel based on its surrounding patches.

Better performance has been achieved using U-Net~\cite{ronneberger2015u}. For example, in~\cite{brosch2016deep}, the authors proposed a 3D-patch-based U-Net that includes an encoder with seven convolutional layers and shortcut connections to integrate multi-scale features.
The performance of a baseline U-Net has been improved using variants of deep neural networks. In~\cite{valverde2017improving}, a cascaded architecture was proposed to enhance the network's sensitivity. In~\cite{hou2019cross}, the authors introduced the attention module in the U-Net framework, improving the model's performance and interpretability. 
Our methods are also based on the U-Net model structure as shown in Figure~\ref{fig:unet}.
% As a comprehensive toolbox, nnU-Net~\cite{isensee2021nnu} is used to automate hyperparameters to tackle domain differences of diverse datasets.

Prior to the application of deep learning models, imaging data requires pre-processing that is intended to normalize the input images by co-registration~\cite{jenkinson2002improved,roura2015multi}, bias correction \cite{tustison2010n4itk}, and skull stripping (\eg, Brain Extraction Tool from FSL~\cite{jenkinson2005bet2}). Where mentioned, these requisite pre-processing have also been employed in our experiments.

%Similarly, we also use a 3D U-Net as our backbone network considering its satisfactory performance on the training data with clean labels.

\subsection{Federated Learning}

The most common usage of federated learning is horizontal federated learning (HFL), in which all sites share the same model structure that deals with different samples with the same data format per site. The trained local site model weights are shared with the central server for aggregation and distribution. Examples of successful applications of deep learning models developed in a federated learning environment include chest x-ray classification for COVID-19~\cite{flores2021federated} and brain tumor segmentation~\cite{sheller2018multi,ng2021federated}. 
% The baseline deep neural networks designed for classification or segmentation can be easily integrated into the federated learning framework, based on open-source toolboxes such as NVIDIA Clara Imaging SDK~\cite{clara}. 
To the best of our knowledge, the current work represents one of the first studies to use horizontal federated learning for the task of MS lesion segmentation.

Current research aims to improve the performance of HFL and deal with more complicated scenarios, including non-independent and non-identically distributed (non-IID) data distributions~\cite{zhao2018federated}, communication efficiency~\cite{konevcny2016federated}, data security and model inversion~\cite{triastcyn2020federated}, and aggregation methods~\cite{li2021fedbn}. Among these studies, the problem of noisy labels, a common issue in real-world applications, has not been considered. Here, we propose a novel training strategy to improve the performance of HFL in the scenario of noisy labels.

\subsection{Noisy Labeling in Segmentation}

% Generating accurate lesion masks from brain MRI images is an error-prone and labor-intensive process that is usually performed on hundreds of 3D FLAIR images per scan. Although a unified lesion identification and labeling process can be provided to participating sites, noisy labels are inevitable due to the heterogeneity of the morphology and appearance of MS lesions under different MRI acquisitions and the variability of different centers including different labeling schemes and bias of annotators. 
%Different with the noisy label problem in the image classification domain that the label of one sample could either be clean or noisy, the label of MRI images 

% Methods for improving the performance of deep learning models trained with noisy labels have been extensively studied, especially in the image classification area.

Recognition of label errors during training by cross-validation can provide an opportunity to remove or correct cases with bad labels or assign lower weights to such cases~\cite{sluban2014ensemble}. Outlier/anomaly detection algorithms can also be used to identify and correct erroneous labels~\cite{han2019deep}. Alternatively, the data labels can be updated with the classifier during the training process~\cite{tanaka2018joint}.
Similarly, segmentation noise can be detected through an additional network~\cite{zhu2019pick, yao2023learning} and data with noisy labels excluded or corrected from the training process. 
% An overfitting control module is proposed to alleviate the overfitting problem introduced by decreasing the size of the dataset.

% teacher-student
Knowledge distillation, in which a teacher network is updated together with a student network, has also proven to be an effective method for dealing with noisy labels. In~\cite{tarvainen2017mean}, the teacher model is updated using the exponential moving average (EMA) of the student model, which helps to mitigate the influence of noisy labels. This framework has been used for segmentation in~\cite{wang2020noise}, where the student network is trained with a noise-robust dice loss, and the teacher model is updated using EMA.
As a special teacher-student network, the co-teaching network~\cite{han2018co} maintains two peer networks that are trained simultaneously to independently select data with clean labels. The knowledge learned from both clean labels is shared between the networks in every iteration. A more complex tri-teaching strategy is proposed in~\cite{zhang2020robust} as an extended co-teaching framework, in which the corrected annotations are selected through consensus and differences between any two of the three teacher networks.

\vspace{1em}

In the current work, we propose a novel label correction strategy, which we refer to as Decoupled Hard Label Correction (DHLC). This strategy builds upon traditional label correction methods~\cite{tanaka2018joint,zhu2019pick,yao2023learning} but focuses on independently adjusting each class with separate hard thresholds. The thresholds are conducted individually for all the voxels and the labels are corrected when only the prediction of the local teacher model and the ground truth labels are contradicted, which are based on the feature of lesions where most of the noise comes from the boundaries.

In the context of federated learning, we utilize the central model aggregated from site models to handle variations in noise distribution across sites. In this Centrally Enhanced Label Correction (CELC) method, the aggregated central model serves as the teacher model for generating pseudo labels. Unlike existing teacher-student network approaches~\cite{tarvainen2017mean, wang2020noise} and in DHLC, the teacher model in CELC is only updated from the aggregation operation in the federated learning central server. This approach helps to eliminate the bias of each local model to the local noise and therefore enhances the label correction strategy within the federated learning framework.

\section{Methodology}
\label{sec:method}

\begin{figure*}[!t]
\centerline{\includegraphics[width=2\columnwidth]{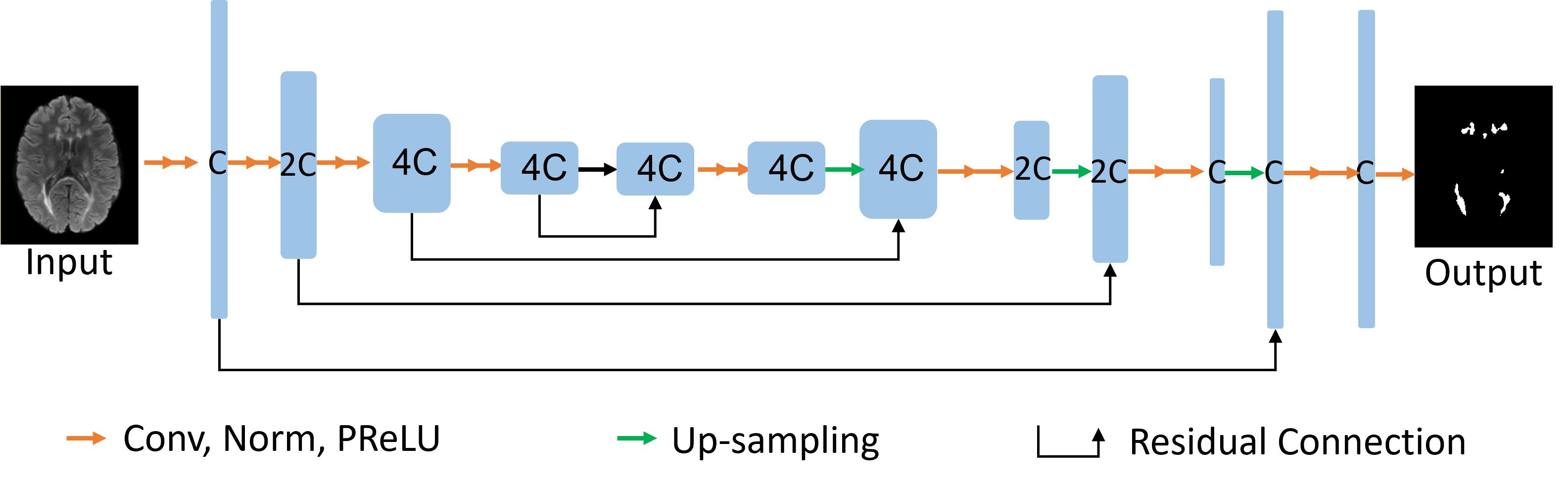}}
\caption{An illustration of the basic U-Net architecture, which is built upon cascade CNN layers with the encoder-decoder structure to capture both low-level contextural patterns and high-level semantic information. Each orange arrow is a convolutional module, including a convolutional layer (``Conv''), a normalization layer (``Norm''), and an activation layer (``PReLU'', which stands for Parametric Rectified Linear Unit). Dual arrows are used to represent the duplicated convolutional modules (refered to as ``dual-conv module''). ``C'' denotes the channels of feature maps, and the dimensions of the feature maps are controlled by strides in convolutional layers and the upsampling layers (marked with green arrows).}
\label{fig:unet}
\end{figure*}

In this section, we first introduce the baseline method for automated lesion segmentation under a multi-site collaborative and distributed learning paradigm. Then, a robust MS lesion segmentation framework (Figure~\ref{fig:framework}), composed of a U-Net-based lesion segmentation network and our new training strategies, is elaborated to ameliorate the impact of noisy labels on the learning process.

\subsection{Lesion Segmentation under Federated Learning}
Our baseline method for lesion segmentation under the federated learning paradigm does not take into account the presence of noisy labels. Following previous work on MS lesion segmentation~\cite{ronneberger2015u}, we utilize the widely used U-Net architecture as the backbone network for mapping input brain MR images to a high-dimensional representation space and for learning salient patterns.

The U-Net architecture is an extension of typical encoder-decoder convolutional neural networks (CNN) designed to capture both low-level contextual patterns and high-level semantic information. As illustrated in Figure~\ref{fig:unet}, the U-Net comprises a down-sampling path and an up-sampling path, each consisting of convolutional modules and dual-conv modules (represented by dual arrows).
The first Conv layers in all the dual-conv modules of the down-sampling path have a stride of 2 (except for the first dual-conv modules, which has a stride of 1 to encode detailed image features), which is intended to reduce the resolution of feature maps and increase the receptive field for subsequent CNN operations. Correspondingly, each dual-conv module in the up-sampling path is followed by an up-convolution operation (a $2 \times 2$ transposed convolution) to expand the resolution of the feature maps. 
Skip connections are employed to preserve more low-level contextual information in the final representation, by concatenating lower-level feature maps with their corresponding higher-level feature maps. Finally, a $1 \times 1$ convolution layer is employed to generate the segmentation map from the final feature map.

Unlike the traditional deep learning training strategy, training the U-Net within the federated learning framework involves multiple \textit{sites} (or local nodes, such as hospitals) and a central \textit{server}  to conduct aggregation of local site models. At the beginning of federated training, each site initializes a local U-Net segmentation model with the same architecture and parameters as the central node. 
%initialized with the same parameters as other sites and updates the model with its local dataset and optimizer. 
During the federated training process, each site optimizes its local model using its site-specific dataset for a certain number of iterations. Subsequently, the site model is uploaded to the central node, where all site models are evenly aggregated to create a more discriminative central model. 
The merged central model is then distributed back to all sites for further local updates. This process of local optimization, model uploading, model aggregation, and model distribution is iteratively performed until the model converges at each site or reaches the predetermined number of local training epochs.

\subsection{Decoupled Hard Label Correction}
The U-Net-based lesion segmentation method described above demonstrates promising performance under the federated learning framework when clean labels are available. 
However, in the presence of noisy labels, additional strategies are needed to train the segmentation model effectively. 
To this end, we introduce a simple, yet effective, strategy that we refer to as Decoupled Hard Label Correction (DHLC), which aims to implicitly identify and correct possible noisy labels at the voxel level during the local optimization stage.

%For an arbitrary input sample $\x_i$ with its corresponding label $\y_i$ at a specific site, a predicted segmentation map generated by U-Net is denoted as $\p_i$, which consists of two channels of the same size as the input. One channel represents the predicted probability for lesions, and the other channel corresponds to the probability for normal tissue.
For an arbitrary input sample $\x_i$ with its corresponding label $\y_i$, a predicted segmentation map generated by U-Net is denoted as $\p_i$. As mentioned in ~\cite{wang2021proselflc}, label correction method (such as~\cite{tanaka2018joint}) mitigates the influence of noisy labels by adjusting the one-hot label distribution $\y_i$ using the predicted distribution $\p_i$:
\begin{equation}
    \label{eq:label_correction}
    \hat{\y}_i = (1 - \epsilon) \y_i + \epsilon \p_i,
\end{equation}
where $\hat{\y}_i$ is the corrected pseudo label for $\x_i$ and $\epsilon$ is a factor to control the correction strength. 
% With the generally used cross entropy loss, the label correction process serves as a regularization method by encouraging the prediction to have lower uncertainty and thus be more meaningful.

% Although the above label correction has shown success in natural image classification, we have observed that it may not be well-suited for the task of MS lesion segmentation. 
% Compared with the natural image classification task, the lesion segmentation task is inherently more ambiguous with relatively fuzzier masks, which lack a clear decision boundary between the lesion and the background. 
This label correction method is intended to generate soft pseudo labels for all samples instead of differentiating between clean labels and noisy labels, which is suitable and has shown success in the natural image classification problems~\cite{tanaka2018joint,yuan2020revisiting}. For the lesion segmentation task which is inherently more ambiguous with relatively fuzzier masks that lack a clear decision boundary between the lesion and background, such a strategy would potentially decrease the model's ability to discriminate between lesion and non-lesion voxels. 
% Therefore, it is more appropriate to apply a voxel-level hard label correction approach that exclusively focuses on explicitly identified noisy labels. 

\begin{figure*}[!t]
\centerline{\includegraphics[width=2\columnwidth]{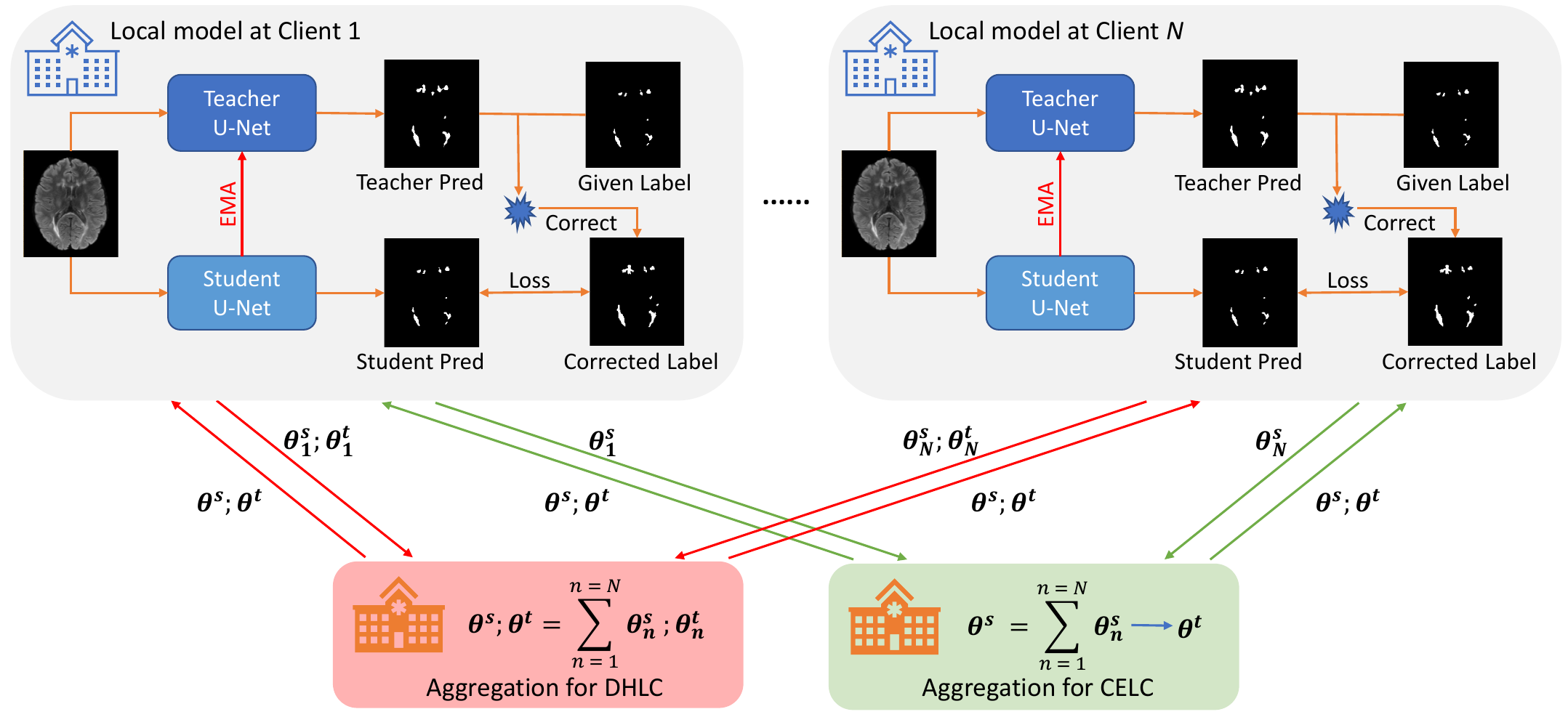}}
\caption{Overall framework for DHLC (red arrows, including EMA) and CELC (green arrows) strategies under the federated learning architecture. $\theta^t$ and $\theta^s$ are the teacher and student networks, respectively. Best viewed in color. More details about CELC are given in Algorithm~\ref{alg}.}
\label{fig:framework}
\end{figure*}

Instead of using the complex tri-teaching strategy~\cite{zhang2020robust} or introducing additional networks~\cite{zhu2019pick} to identify noisy labels, we propose to identify and correct noisy labels at the voxel level based on the following rules (as shown in Equation~\ref{eq:core}): 1) if voxel $x_j$ is predicted as not being a lesion with a probability higher than a threshold $H_0$ but the given label identifies it is a lesion, the label is then considered noisy and should be corrected as not a lesion (\ie, the background); 2) if voxel $x_j$ is predicted as a lesion with a probability higher than a threshold $H_1$ but the given label does not identify it as a lesion, then the label is also considered noisy and should be corrected as a lesion; 
3) otherwise, the original label is preserved.
\begin{equation}
   \label{eq:core}
   \hat{y}_{j}= 
   \begin{cases} 
   0 & \text{if} \arg \max _{k} \p_{j}^{k}=0, \p_{j}^{0}>H_0, y_j=1,
\\ 1 & \text{if} \arg \max _{k} \p_{j}^{k}=1, \p_{j}^{1}>H_1, y_j=0, 
\\ y_{j} & \text{otherwise,}
   \end{cases} 
\end{equation}
where $y_j$ and $\hat{y}_j$ are the original label and the updated label for voxel $x_j$, $\p_j$ is the prediction vector of the voxel $x_j$. In our strategy, the $\p_j$ vector is generated by the local teacher network with the weights $\bm{\theta}^t_n$, which is updated based on the local student network weights $\bm{\theta}^s_n$ using the EMA method mentioned in~\cite{tarvainen2017mean,wang2020noise}.

% Note that $H_0$ and $H_1$ are typically not equal due to the highly imbalanced nature of the MS lesion dataset, where lesions occupy only a very small portion of the brain.
%Here, we only correct the noisy labeled voxels to lesion (positive) and name the method as positive label correction considering the practical annotation process in MS lesion area. Due to the fuzzy nature of MS lesions, the data labeling process is rather time-consuming and the divergence across annotator are high. Thus, utilizing a conservative data labeling protocol could both save the annotation time and decrease the divergence. In such case, give labels are normally smaller than the ground-truth labels and only positive label correction is enough. If omitting this practical annotation protocol, the label correction should be applied to both positive and negative classes as.
%for two reasons. First, the segmentation mask of existing deep learning models are still not enough accurate, correcting noisy labels to background (negative)   good the lesion segmentation task is extremely imbalanced, where most voxels in the input are normal tissue. 

\subsection{Centrally Enhanced Label Correction}
While the hard label correction significantly improves the model's robustness against noisy labels, we have observed that the aggregated central model, which is more discriminative, tends to be deteriorated by local models that are trained with heavy label noise. This is inherently caused by the FedAvg algorithm~\cite{mcmahan2017communication}, in which multiple optimization iterations are performed at each site before aggregation and the site models tend to overfit to the label noise, thereby affecting the performance of the aggregated central model and site models, and the robustness of the model when trained to convergence.

To mitigate this issue, we propose using the central model as the teacher model to guide the learning of the student segmentation network (\ie, local models from sites), as shown in Figure~\ref{fig:framework}.
This design is motivated by the observation that the central model, which is aggregated from multiple sites, is less prone to overfitting to local label noise compared to the site models that are trained with heavily corrupted labels. Thus, using the central model to correct the label noise is likely to yield more accurate results.

\begin{algorithm}
\caption{Training and Optimization Pipeline of CELC} 
\label{alg}
\begin{algorithmic}[1]
\REQUIRE $N$ sites $\mathcal{S} = \{(S_1, S_2, ..., S_N\}$ that participate in federated learning, each with a labeled (unknown noisy or clean) training dataset $\mathcal{D}_n = \{(\x, \y)\}$ ($n \in [1, N]$), maximum aggregation rounds $R$, training iteration $I_{n}$ at each site $n$ per round, and a clean labeled validation set in the center $\mathcal{D}_c = \{(\x_c, \y_c)\}$. \{$\bm{\theta}^t$, $\bm{\theta}^s$\} are the parameters of the teacher model and student model.
\STATE \{$\bm{\theta}^t$, $\bm{\theta}^s$\} = RandomInitialize() in the center and assign them to all sites
\FOR{$\text{round }r=0$; $r<R$}
\FOR{$\text{site }n=0$; $n<N$}
\STATE Retrieve the site model, optimizer, data loader
\FOR{$iter=0; iter<I_{n}$}
\STATE Sample a batch $\{(\x_i, \y_i)\}_{iter}$ from $\mathcal{D}_{n}$
\STATE Feed $\{\x_i\}_{iter}$ to the model and get the prediction scores $\{\p_i\}_{iter}$
\STATE Get the corrected label $\{\hat{\y}_i\}_{iter}$ with Equation~\ref{eq:core}
\STATE Calculate the cross entropy loss with $\{\hat{\y}_i\}_{iter}$
\STATE Backpropagate and update $\bm{\theta}_{n}^{s}$
\ENDFOR
\STATE Upload the site model weights $\{\bm{\theta}_{n}^{s}\}$ to the center
\ENDFOR
\STATE Merge weights $\{\bm{\theta}_{n}^{s}\}$ to obtain central model $\bm{\theta}^s$
\STATE Update $\bm{\theta}^t$ with the best central model selected with the center validation set $\mathcal{D}_c$ 
\STATE Distribute the central model \{$\bm{\theta}^t$, $\bm{\theta}^s$\} to each site
\ENDFOR
\end{algorithmic}
\end{algorithm}

The algorithm process of this Centrally Enhanced Label Correction (CELC) strategy is shown in~\ref{alg}. The correction strategy of CELC is the same as that of DHLC in~\ref{eq:core}, where $\q_{j}$ is the predicted probability distribution of voxel $x_j$ from the teacher model with weights $\bm{\theta}^t$ obtained from the aggregation server. 
In contrast to the teacher-student network design presented in~\cite{wang2020noise}, the teacher model in CELC is not updated during the site optimization process. Rather, it is updated by the center as part of the federated learning framework (as shown in Figure~\ref{fig:framework}). Specifically, when the newly aggregated model performs worse than the previous best central model in the validation dataset, the previous best central model is used as the teacher model and shared with sites in the next round.

% \begin{equation}
%    \hat{y}_{j}= 
%    \begin{cases} 
%    0 & \text{if} \arg \max _{k} \hat{\q}_{j}^{k}=0, \hat{\q}_{j}^{0}>H_0, y_j=1,
%    \\ 1 & \text{if} \arg \max _{k} \hat{\q}_{j}^{k}=1, \hat{\q}_{j}^{1}>H_1, y_j=0, 
%    \\ y_{j} & \text {otherwise,}
%    \end{cases} 
%    \label{eq:celc}
% \end{equation}

\section{Experiments}
\label{sec:experiments}
In this section, we present the empirical analysis of the influence of noisy labels on MS lesion segmentation and the performance of our methods. 

% Other potential strategies exist for leveraging the central model information to address noisy labels. One such strategy is using the central model as a teacher to guide the optimization process of each site model through knowledge distillation, in addition to label correction. This approach provides additional regularization and helps reduce the risk of overfitting at heavily corrupted sites. However, in our empirical analysis, we observed that the proposed center-enhanced label correction strategy already achieves performance similar to that of the model trained with clean labels (as shown in Section~\ref{sec:performance}). Therefore, introducing further regularization through knowledge distillation did not lead to an improvement in performance.

\subsection{Experimental Settings}
% The experimental settings are as follows, including the datasets used, implementation details, and evaluation metrics.
\subsubsection{Datasets} 
Two multi-site datasets are employed in our experiments.

The first dataset is MSSEG-2016\cite{commowick2021multiple}, which was used in the MICCAI MS lesion segmentation challenge in 2016. It consists of 53 MRI images from four different data centers, with labels determined through the consensus of seven expert annotators. The original training data contains 15 cases from three data centers, while the test set contains 38 MRI images, including 30 samples from the training centers and extra 8 cases from an additional center. 
As the domain difference is not the key aspect of our paper, we excluded the 8 cases from the additional center and followed the original settings of the remaining three sites by assigning 5 training samples to each site. In contrast with the original data split, 2 samples from each original test set were extracted to form the validation set, and the remaining 24 samples were used for testing and reporting the results. 
To simulate varying levels of label corruption across different sites, different strategies were adopted to introduce noise to the training labels. The original clean labels were retained for the first site. For the second and third sites, we introduced corruptions by eroding or dilating 40\% of the training samples, respectively, with two voxels in the lesion mask (referred to as ``Label Erosion'' or ``Label Dilation'') to mimic the noise caused by inconsistent mask boundaries. Given that MS lesions are typically small in size, dilation or erosion by two voxels can be considered a reasonable form of noise.

The second dataset (SNAC-MS) comprises 123 cases collected internally from two sites, with labels annotated by two professional neuroimaging analysts in parallel, followed by cross-checking by another expert annotator. To avoid the influence of domain differences and non-independent and non-identical distributions across centers, the dataset was divided into four sites regardless of the original sites of the collection, with each site containing 20 samples for training. We reserved 21 cases for testing, and the remaining 22 samples were used as the center validation set.
Similar to the first dataset, we retained the clean labels for the first site. For the second and third sites, we applied Label Erosion to 40\% or 80\% of the training labels. The training labels in the fourth site were corrupted by randomly removing 40\% of the lesions (referred to as ``Label Removal'') to mimic the noise caused by conservative annotators. The noise related to additional false positive lesions was not considered as false positive lesions are rarely labeled by human annotators according to our experience and the previous research~\cite{whiting2006accuracy} in which the specificity is pretty high in diagnosis when MRI images are presented. 

%, except that lesion re Similar to the corruption strategy applied to SNAC-MS, we keep the clean label for the first site and corrupt the label of the second and third site. In detail, labels of 40\% samples in the second site are corrupted by the erosion corruptions in the lesion mask with two voxels and 

All samples in both datasets consist of co-registered FLAIR and T1 images as input. Skull stripping was performed on these images, and all generated brain images were registered to the MNI space~\cite{grabner2006symmetric} using FLIRT~\cite{greve2009accurate} for pre-processing and standardization purposes. The performance was reported on the test set with the same pre-processing steps using clean labels.

\subsubsection{Implementation Details}
For federated learning, all sites utilized the same hyperparameter settings.
Inspired by prior studies~\cite{wang2020noise}, we adopted a 2D U-Net as the segmentation network, processing slices individually from the original 3D MRI images to accommodate the large size of each FLAIR and T1 image.
The U-Net has a depth of 3, \ie, three consecutive dual-conv modules were incorporated for encoder and decoder. The unit number of channel is set to 32 (\ie, the ``C'' shown in Figure~\ref{fig:framework}). 
The network was optimized with the Adam optimizer with a learning rate of 9.0e-4 for the MSSEG-2016 and 7.0e-4 for the SNAC-MS dataset, with weight decay of 1.0e-5. The batch size was set to 32 and the total optimization step was 50 epochs with models aggregated after each epoch. Regarding the label correction strategy, $\{H_0, H_1\}$ was set to $\{0.90, 0.65\}$ for the MSSEG-2016 dataset and $\{1, 0.65\}$ for the SNAC-MS dataset, which were decided upon the validation datasets. Prior to applying the label correction strategy, 10 warm-up pre-train epochs were conducted in which the label corrections were not conducted to ensure that the generated pseudo labels were of reasonable quality.

\begin{table*}[]
\centering
\caption{The influence of noisy labels to both centralized training and federated training paradigms.}
\label{tab:influence}
\begin{tabular}{c|c|cccccccc}
\hline
\multirow{4}{*}{\begin{tabular}[c]{@{}c@{}}Learning \\ Paradigm\end{tabular}} & 
\multirow{4}{*}{\begin{tabular}[c]{@{}c@{}}Label\\ Noise\end{tabular}} & 
\multicolumn{8}{c}{Datasets} \\ \cline{3-10}   &
& \multicolumn{4}{c|}{MSSEG-2016}
& \multicolumn{4}{c}{SNAC-MS} \\ \cline{3-10} & 
& \multicolumn{1}{c|}{P-Dice $\uparrow$} 
& \multicolumn{1}{c|}{V-Dice $\uparrow$} 
& \multicolumn{1}{c|}{Precision $\uparrow$} 
& \multicolumn{1}{c|}{Recall $\uparrow$} 
& \multicolumn{1}{c|}{P-Dice $\uparrow$} 
& \multicolumn{1}{c|}{V-Dice $\uparrow$} 
& \multicolumn{1}{c|}{Precision $\uparrow$} 
& \multicolumn{1}{c}{Recall $\uparrow$} \\ \hline
Centralized & No
% \multicolumn{1}{c|}{0.4984} & \multicolumn{1}{c|}{0.6437} & \multicolumn{1}{c|}{0.5052} & 0.6516 \\ \hline
& \multicolumn{1}{c|}{0.6273} & \multicolumn{1}{c|}{0.7396} & \multicolumn{1}{c|}{0.6570} & \multicolumn{1}{c|}{0.6361}
& \multicolumn{1}{c|}{0.7066} & \multicolumn{1}{c|}{0.7725} & \multicolumn{1}{c|}{0.7220} & \multicolumn{1}{c}{0.7057} 
\\ \hline
Centralized & YES
% \multicolumn{1}{c|}{0.4717} & \multicolumn{1}{c|}{0.5303} & \multicolumn{1}{c|}{0.6519} & 0.4300 \\ \hline
& \multicolumn{1}{c|}{0.5886} & \multicolumn{1}{c|}{0.6904} & \multicolumn{1}{c|}{0.6959} & \multicolumn{1}{c|}{0.5539}
& \multicolumn{1}{c|}{0.6698} & \multicolumn{1}{c|}{0.7206} & \multicolumn{1}{c|}{0.7682} & \multicolumn{1}{c}{0.6092}
\\ \hline
Federated & No 
& \multicolumn{1}{c|}{0.6108} & \multicolumn{1}{c|}{0.7253} & \multicolumn{1}{c|}{0.6813} & \multicolumn{1}{c|}{0.6335} 
& \multicolumn{1}{c|}{0.7097} & \multicolumn{1}{c|}{0.7688} & \multicolumn{1}{c|}{0.7391} & \multicolumn{1}{c}{0.6904}
% \multicolumn{1}{c|}{0.5155} & \multicolumn{1}{c|}{0.6221} & \multicolumn{1}{c|}{0.5307} & 0.6443 \\ \hline
\\ \hline
Federated & YES
% \multicolumn{1}{c|}{0.4907} & \multicolumn{1}{c|}{0.5464} & \multicolumn{1}{c|}{0.6263} & 0.4815 \\ \hline
& \multicolumn{1}{c|}{0.5157} & \multicolumn{1}{c|}{0.6530} & \multicolumn{1}{c|}{0.5999} & \multicolumn{1}{c|}{0.5004}
& \multicolumn{1}{c|}{0.6213} & \multicolumn{1}{c|}{0.6944} & \multicolumn{1}{c|}{0.8366} & \multicolumn{1}{c}{0.5006} 
\\ \hline
\end{tabular}
\end{table*}

\subsubsection{Evaluation Metrics}
To quantitatively evaluate and compare the performance of different methods, we employed four evaluation metrics: subject-level Dice Similarity (P-Dice), voxel-level Dice Similarity (V-Dice), and subject-level average Precision and Recall.

\begin{equation}
\begin{aligned}
\text{P-Dice} (\p, \y) &= \frac {2}{T} \times \sum \frac{|\p_i \cap \y_i|}{|\p_i| + |\y_i|} \\
\text{V-Dice } (\p, \y) &= \frac{2 \times \sum |\p_i \cap \y_i|}{\sum |\p_i| + \sum |\y_i|} \\
% Dice (\p, \y) &= \frac{2 * \sum (\p \cap \y)}{\p + \y} \\
\text{Precision } (\p, \y) &= \frac {1}{T} \times \sum \frac{|\p_i \cap \y_i| }{|\p_i|} \\
\text{Recall } (\p, \y) &= \frac {1}{T} \times \sum \frac{|\p_i \cap \y_i|}{|\y_i|},
\end{aligned}
\end{equation}
where $\p$ is the predicted lesion map, $\y$ is the ground truth lesion map, and $T$ is the total number of subjects. $|\cdot|$ denotes the sum of the elements within the tensor, and $\sum$'s are all conducted across all test cases. 
P-Dice and V-Dice are selected as the main figure to report performance. In addition, subject-level Precision and Recall are used to reflect the methods' performance on subjects, regardless of individual lesion volumes.

\subsection{Influence of Noisy Labels} \label{sec:influence}
As there is no prior work on federated MS lesion segmentation with noisy labels, we begin by analyzing the influence of noisy labels on both the centralized learning paradigm and the federated learning paradigm. Under the centralized learning paradigm, the training data from different sites are merged for training, while the validation and test set remain the same as in the federated learning paradigm. The experimental results are presented in Table~\ref{tab:influence}.

For the MSSEG-2016 dataset, federated training demonstrates similar performance to the centralized training scheme in the absence of label noise, demonstrating the feasibility of learning discriminative MS lesion segmentation models without data sharing across sites to maintain privacy.
However, in the presence of label noise, the performance of the federated learning paradigm deteriorated more than the centralized training. Specifically, centralized training achieves a P-Dice of 0.5886 and a V-Dice of 0.6904, which are 0.0387 and 0.0492 lower than the centralized model trained with clean labels. In contrast, federated training with noisy labels only achieves a P-Dice of 0.5157 and a V-Dice of 0.6530, showing decreases of 0.0951 and 0.0723 compared with the federated model trained with clean labels. 
This deterioration can be attributed to the inherent problem in federated learning, as previously described in~\cite{mcmahan2017communication}. The site models are more prone to overfitting the label noise during their individual updates, which subsequently affects the performance of the averaged model and leads to further performance degradation.
Moreover, it is evident that the introduced label corruption significantly decreases the recall in both centralized and federated training, suggesting that many lesions are overlooked by the model. However, the precision of the centralized model, when trained with noisy labels, shows an improvement compared to centralized training with clean labels. This observation indicates that the combined effects of dilation and erosion noise may have balanced out in the centralized training. However, this phenomenon is not observed in the federated learning paradigm.

For the SNAC-MS dataset, the performance also drops when there are label corruptions for both centralized training and federated training. Unlike the observation in the MSSEG-2016 dataset, both training paradigms exhibit improved precision with a sacrifice in recall. This phenomenon may be related to the specific label corruption operations applied in this dataset. Both Label Erosion and Label Removal tend to generate false negatives, leading to the bias of the model on the eroded lesion masks. Such an effect is more obvious in the federated learning paradigm, further demonstrating that it is more susceptible to be overfitting on site-specific label noise under standard federated learning settings. 

\subsection{Effectiveness of CELC Strategy}

\begin{table*}[]
\centering
\caption{Overall comparison of the proposed method and existing noisy labeling models for combating label noise. }
\label{tab:comparison}
\begin{tabular}{c|cccccccc}
\hline
\multirow{4}{*}{Method} 
& \multicolumn{8}{c}{Datasets} \\ \cline{2-9} 
& \multicolumn{4}{c|}{MSSEG-2016}
& \multicolumn{4}{c}{SNAC-MS} \\ \cline{2-9} 
& \multicolumn{1}{c|}{P-Dice $\uparrow$} 
& \multicolumn{1}{c|}{V-Dice $\uparrow$} 
& \multicolumn{1}{c|}{Precision $\uparrow$} 
& \multicolumn{1}{c|}{Recall $\uparrow$} 
& \multicolumn{1}{c|}{P-Dice $\uparrow$} 
& \multicolumn{1}{c|}{V-Dice $\uparrow$} 
& \multicolumn{1}{c|}{Precision $\uparrow$} & Recall $\uparrow$\\ \hline
Basic U-Net (Clean) 
& \multicolumn{1}{c|}{0.6108} & \multicolumn{1}{c|}{0.7253} & \multicolumn{1}{c|}{0.6813} & \multicolumn{1}{c|}{0.6335} 
& \multicolumn{1}{c|}{0.7097} & \multicolumn{1}{c|}{0.7688} & \multicolumn{1}{c|}{0.7391} & \multicolumn{1}{c}{0.6904}
\\ \specialrule{2pt}{0pt}{0pt}

Basic U-Net 
& \multicolumn{1}{c|}{0.5157} & \multicolumn{1}{c|}{0.6530} & \multicolumn{1}{c|}{0.5999} & \multicolumn{1}{c|}{0.5004}
& \multicolumn{1}{c|}{0.6213} & \multicolumn{1}{c|}{0.6944} & \multicolumn{1}{c|}{0.8366} & \multicolumn{1}{c}{0.5006} 
\\ \hline
Label Smoothing~\cite{muller2019does}
% \multicolumn{1}{c|}{0.5163} & \multicolumn{1}{c|}{0.5838} & \multicolumn{1}{c|}{0.6439} &   0.4974  \\ \hline
& \multicolumn{1}{c|}{0.5187} & \multicolumn{1}{c|}{0.6580} & \multicolumn{1}{c|}{0.5505} & \multicolumn{1}{c|}{0.5589} 
& \multicolumn{1}{c|}{0.6026} & \multicolumn{1}{c|}{0.6838} & \multicolumn{1}{c|}{0.8204} & \multicolumn{1}{c}{0.4872} 
\\ \hline
Label Correction~\cite{wang2021proselflc}
% \multicolumn{1}{c|}{0.4943} & \multicolumn{1}{c|}{0.5318} & \multicolumn{1}{c|}{0.6648} &  0.4621 \\ \hline
& \multicolumn{1}{c|}{0.5777} & \multicolumn{1}{c|}{0.6948} & \multicolumn{1}{c|}{\textbf{0.7074}} & \multicolumn{1}{c|}{0.5380}
& \multicolumn{1}{c|}{0.6010} & \multicolumn{1}{c|}{0.6832} & \multicolumn{1}{c|}{\textbf{0.8438}} & \multicolumn{1}{c}{0.4798} 
\\ \hline
ProSelfLC~\cite{wang2021proselflc}
% \multicolumn{1}{c|}{0.3454} & \multicolumn{1}{c|}{0.4937} & \multicolumn{1}{c|}{0.3685} & \multicolumn{1}{c}{0.5074} \\ \hline
& \multicolumn{1}{c|}{0.5648} & \multicolumn{1}{c|}{0.6955} & \multicolumn{1}{c|}{0.6938} & \multicolumn{1}{c|}{0.5429}
& \multicolumn{1}{c|}{0.6397} & \multicolumn{1}{c|}{0.7097} & \multicolumn{1}{c|}{0.8355} & \multicolumn{1}{c}{0.5325} 
\\ \hline
% Pick\&Learn & 
% \multicolumn{1}{l|}{}       & \multicolumn{1}{l|}{}       & \multicolumn{1}{l|}{}          & \multicolumn{1}{l|}{}       & \multicolumn{1}{l|}{}       & \multicolumn{1}{l|}{}       & \multicolumn{1}{l|}{}          & \multicolumn{1}{l}{} \\ \hline
PINT~\cite{shi2021distilling}
% \multicolumn{1}{c|}{0.5167} & \multicolumn{1}{c|}{0.5572} & \multicolumn{1}{c|}{0.6580} & \multicolumn{1}{c}{0.5021} \\ \hline
& \multicolumn{1}{c|}{0.5576} & \multicolumn{1}{c|}{0.6871} & \multicolumn{1}{c|}{0.6521} & \multicolumn{1}{c|}{0.5550} 
& \multicolumn{1}{c|}{0.6450} & \multicolumn{1}{c|}{0.7094} & \multicolumn{1}{c|}{0.8370} & \multicolumn{1}{c}{0.5375} 
\\ \specialrule{2pt}{0pt}{0pt}
U-Net+DHLC (Ours)
& \multicolumn{1}{c|}{0.5898} & \multicolumn{1}{c|}{0.6943} & \multicolumn{1}{c|}{0.6542} & \multicolumn{1}{c|}{0.6354}
& \multicolumn{1}{c|}{0.6994} & \multicolumn{1}{c|}{0.7518} & \multicolumn{1}{c|}{0.6992} & \multicolumn{1}{c}{0.7063}
\\ \hline
U-Net+CELC (Ours)
% \multicolumn{1}{c|}{0.5272} & \multicolumn{1}{c|}{0.6521} & \multicolumn{1}{c|}{0.5620} & 0.6155 \\ \hline
& \multicolumn{1}{c|}{\textbf{0.6082}} & \multicolumn{1}{c|}{\textbf{0.7156}} & \multicolumn{1}{c|}{0.6431} & \multicolumn{1}{c|}{\textbf{0.6822}}
& \multicolumn{1}{c|}{\textbf{0.7034}} & \multicolumn{1}{c|}{\textbf{0.7655}} & \multicolumn{1}{c|}{0.7433} & \multicolumn{1}{c}{\textbf{0.7175}} 
\\ \hline
\end{tabular}
\end{table*}

\label{sec:performance}
In this section, we evaluate and report the performance of our method (CELC) in handling noisy labels on both the MSSEG-2016 and SNAC-MS datasets. To provide a comprehensive evaluation, we compare our method with four representative approaches in the noisy labeling area:

\begin{itemize}
    \item Label Smoothing~\cite{muller2019does} is a baseline strategy for noisy labeling in the natural image classification area, which softens the one-hot class label with a weighted constant during the training process to decrease over-fitting. 
    \item Label Correction~\cite{wang2021proselflc} is a similar way to Label Smoothing, except that it uses the weighted combination of the predicted label and the original label as the learning target. 
    \item ProSelfLC~\cite{wang2021proselflc} builds upon the Label Correction method, which progressively updates the weights of the prediction in the learning target by considering both the learning time (\ie, passed iterations divided by total iterations) and the prediction entropy.
    %\item Pick\&Learn~\cite{zhu2019pick} is a noisy labeling method for medical image segmentation tasks, which employs an auxiliary network to learn the quality of the provided segmentation label together with the optimization of the segmentation network (e.g., the U-Net). 
    \item PINT~\cite{shi2021distilling} is a recent noisy labeling method for medical image segmentation, which estimates both the voxel-level label quality and image-level label quality by measuring the prediction uncertainties with an auxiliary network.
\end{itemize}

The experimental results of our model and these comparison methods are presented in Table~\ref{tab:comparison}. As observed, Label Smoothing~\cite{muller2019does}, Label Correction~\cite{wang2021proselflc}, ProSelfLC~\cite{wang2021proselflc}, and PINT~\cite{shi2021distilling} do not consistently boost the performance with noisy-labeled training data for both datasets, which reveals the challenge of training with noisy labeled data under the federated learning paradigm and the difficulty of adapting existing noisy labeling methods developed for the natural image classification tasks to medical image segmentation scenarios. 
Specifically, the Label Correction method achieved the highest precision scores in both datasets, but recall was largely sacrificed, leading to a suboptimal P-Dice and V-Dice scores.
Nevertheless, our method CELC outperformed all methods and significantly improved the performance of the segmentation network, \ie, 0.0925 and 0.0626 performance gain on the MSSEG-2016 dataset and 0.0821 and 0.0711 performance gain on the SNAC-MS dataset in terms of the P-Dice and V-Dice scores. 

By comparing the results in Table~\ref{tab:influence}, we observe that our method CELC achieves a performance close to the model trained with clean labels under the federated learning scheme on both datasets, which demonstrates that our CELC strategy is robust to corrupted labels.

\begin{figure}[!t]
\centerline{\includegraphics[width=\columnwidth]{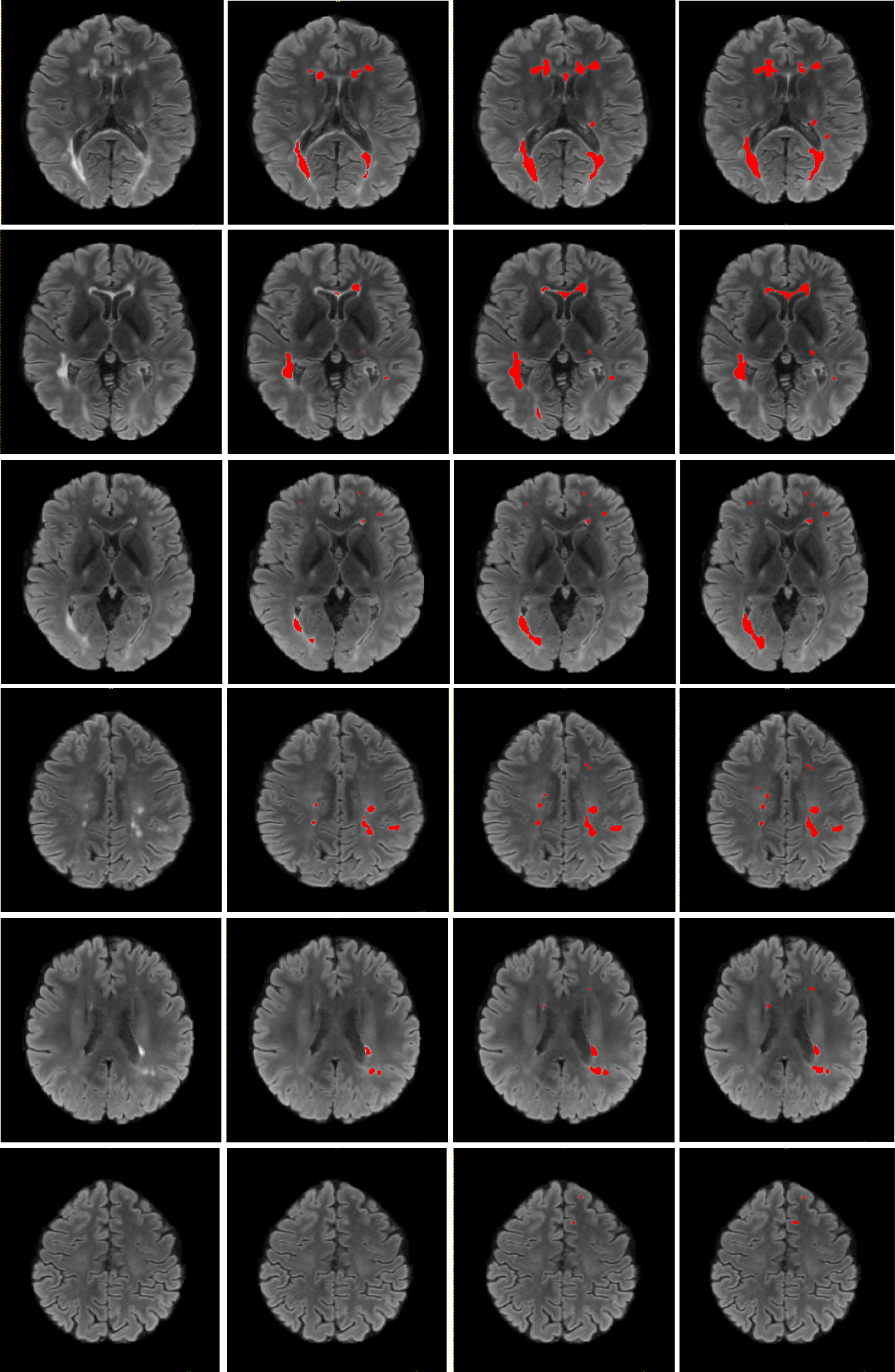}}
\caption{Visualization of different MRI slices, where images on 1) column one are input MRI slices (in FLAIR), 2) column two are segmentation results of the basic U-Net model, 3) column three are segmentation results of our method, and 4) column four are the ground-truth label. Best be viewed by zooming in for small lesions, \eg, the fourth to sixth rows.}
\label{fig:visualization}
\end{figure}

Figure~\ref{fig:visualization} presents an example visualization of the segmentation results obtained by the basic U-Net model and our method, along with the provided ground-truth label from the SNAC-MS dataset.
This visualization reveals that our method achieves reliable segmentation performance despite the presence of severe label noise in the training data. Especially, our method CELC consistently identifies more accurate lesion regions, while the basic U-Net model tends to make incorrect predictions on the lesion boundary and ignore small lesion regions. 

\subsection{Compare CELC and DHLC}
To assess the individual contributions of the decoupled hard label correction (DHLC) and the center-enhanced label correction (CELC) strategy to the observed performance improvements, we further conducted a comparison study of DHLC and CELC as presented in Table~\ref{tab:comparison}. 
The variant ``U-Net+DHLC'' incorporates the hard label correction strategy during the training of the U-Net segmentation model, focusing solely on the site level. As can be observed, ``U-Net+DHLC'' already achieves promising performance by mitigating the influence of noisy labels on network optimization and preventing the overfitting issues associated with federated learning using noisy labels, as discussed in Section~\ref{sec:influence}. 
However, leveraging global knowledge of the federated learning paradigm further enhances the model's robustness. By utilizing the best central model for label correction, our method (\ie, ``U-Net+CELC'') achieves additional performance improvements of 0.0184 and 0.0213 on the MSSEG-2016 dataset and 0.004 and 0.0137 on the SNAC-MS dataset in terms of the P-Dice and V-Dice scores when compared to ``U-Net+DHLC''.

% \begin{table*}[]
% \centering
% \caption{Ablation study}
% \label{tab:ablation}
% \begin{tabular}{c|cccccc}
% \hline
% \multirow{3}{*}{Method} & \multicolumn{6}{c}{Datasets} \\ \cline{2-7} 
%  & \multicolumn{3}{c|}{MSSEG-2016} 
%  & \multicolumn{3}{c}{SNAC-MS} \\ \cline{2-7} 
%  & \multicolumn{1}{c|}{V-Dice $\uparrow$} 
%  & \multicolumn{1}{c|}{Precision $\uparrow$} 
%  & \multicolumn{1}{c|}{Recall $\uparrow$} 
%  & \multicolumn{1}{c|}{V-Dice $\uparrow$} 
%  & \multicolumn{1}{c|}{Precision $\uparrow$} 
%  & Recall $\uparrow$ \\ \hline
% Basic U-Net 
% & \multicolumn{1}{c|}{0.6080} & \multicolumn{1}{c|}{0.6108} & \multicolumn{1}{c|}{0.4416}
% & \multicolumn{1}{c|}{0.6944} & \multicolumn{1}{c|}{0.8366} & \multicolumn{1}{c}{0.5006}
% \\ \hline
% U-Net+DHLC 
% & \multicolumn{1}{c|}{0.6817} & \multicolumn{1}{c|}{0.6412} & \multicolumn{1}{c|}{0.5695}
% & \multicolumn{1}{c|}{0.7518} & \multicolumn{1}{c|}{0.6992} & \multicolumn{1}{c}{0.7063}
% \\ \hline
% U-Net+CELC 
% & \multicolumn{1}{c|}{0.7081} & \multicolumn{1}{c|}{0.6702} & \multicolumn{1}{c|}{0.5818}
% & \multicolumn{1}{c|}{0.7655} & \multicolumn{1}{c|}{0.7433} & \multicolumn{1}{c}{0.7175}
% \\ \hline
% \end{tabular}
% \end{table*}

\subsection{Analysis}
\subsubsection{Influence of the Correction Threshold}
The correction threshold is a critical hyperparameter in our method as it determines whether a given voxel label should be corrected. In this section, we analyze the influence of the correction threshold by varying its value for the training process. Here, we present the analysis using the threshold $H_1$ on the SNAC-MS dataset as an example while reserving $H_0=1.0$ as the control variable.
As shown in Table~\ref{tab:threshold}, our method achieves resilient performance when the threshold is set between 0.55 and 0.75. This demonstrates the robustness of our method to hyperparameters and makes the method more readily applicable in other scenarios. When the threshold is too large (\eg, above 0.8 for $H_1$), the performance noticeably drops as the model fails to correct a sufficient number of false annotations.

\begin{table}[]
\centering
\caption{Segmentation performance with different positive label correction threshold ($H_1$) on the SNAC-MS dataset using CELC.}
\label{tab:threshold}
\begin{tabular}{c|cccc}
\hline
Parameter & \multicolumn{4}{c}{Metrics}   \\ \hline
$H_1$ 
& \multicolumn{1}{c|}{P-Dice $\uparrow$} 
& \multicolumn{1}{c|}{V-Dice $\uparrow$} 
& \multicolumn{1}{c|}{Precision $\uparrow$} 
& \multicolumn{1}{c}{Recall $\uparrow$}  \\ \hline
0.55      
& \multicolumn{1}{c|}{0.6948} & \multicolumn{1}{c|}{0.7596} & \multicolumn{1}{c|}{0.6848}  & 0.7665 \\ \hline
0.60      
& \multicolumn{1}{c|}{0.7020} & \multicolumn{1}{c|}{0.7645} & \multicolumn{1}{c|}{0.7072}  & 0.7537 \\ \hline
0.65      
& \multicolumn{1}{c|}{0.7034} & \multicolumn{1}{c|}{0.7655} & \multicolumn{1}{c|}{0.7433}  & 0.7175 \\ \hline
0.70     
& \multicolumn{1}{c|}{0.6966} & \multicolumn{1}{c|}{0.7626} & \multicolumn{1}{c|}{0.7398}  & 0.6995 \\ \hline
0.75    
& \multicolumn{1}{c|}{0.6845} & \multicolumn{1}{c|}{0.7547} & \multicolumn{1}{c|}{0.7819}  & 0.6524 \\ \hline
0.80   
& \multicolumn{1}{c|}{0.6554} & \multicolumn{1}{c|}{0.7270} & \multicolumn{1}{c|}{0.8375}  & 0.5775 \\ \hline
\end{tabular}
\end{table}

\subsubsection{Influence of the Warm-up Epoch}
We also analyzed the influence of another hyperparameter in our method, namely the number of warm-up epochs, and present the experimental results in Table~\ref{tab:warm-up}.
The model consistently achieves good performance with different numbers of warm-up epochs as long as the warm-up is employed. This demonstrates the robustness of our method regardless of the stage at which it is applied during training.

\begin{table}[]
\centering
\caption{Lesion segmentation performance of CELC with different warm-up epochs on the SNAC-MS dataset.}
\label{tab:warm-up}
\begin{tabular}{c|cccc}
\hline
Parameter & \multicolumn{4}{c}{Metrics}  \\ \hline
Warm-up 
& \multicolumn{1}{c|}{P-Dice $\uparrow$} 
& \multicolumn{1}{c|}{V-Dice $\uparrow$} 
& \multicolumn{1}{c|}{Precision $\uparrow$} 
& \multicolumn{1}{c}{Recall $\uparrow$}  \\ \hline
0 
& \multicolumn{1}{c|}{0.2138} & \multicolumn{1}{c|}{0.2585} & \multicolumn{1}{c|}{0.1363}  & 0.7393 \\ \hline
2     
& \multicolumn{1}{c|}{0.7007} & \multicolumn{1}{c|}{0.7649} & \multicolumn{1}{c|}{0.7317}  & 0.7052 \\ \hline
5     
& \multicolumn{1}{c|}{0.7023} & \multicolumn{1}{c|}{0.7648} & \multicolumn{1}{c|}{0.7208}  & 0.7189 \\ \hline
10     
& \multicolumn{1}{c|}{0.7033} & \multicolumn{1}{c|}{0.7655} & \multicolumn{1}{c|}{0.7433}  & 0.7175 \\ \hline
15     
& \multicolumn{1}{c|}{0.6983} & \multicolumn{1}{c|}{0.7635} & \multicolumn{1}{c|}{0.7054}  & 0.7153 \\ \hline
20     
& \multicolumn{1}{c|}{0.7005} & \multicolumn{1}{c|}{0.7642} & \multicolumn{1}{c|}{0.7177}  & 0.7086 \\ \hline
\end{tabular}
\end{table}

%\subsubsection{Learning the Site Label Quality}
%Except for enhancing the label correction effectiveness, the teacher model we introduced could also help learn the relative site label quality of all sites attending the federated learning process. Since the teacher model is fixed after

\subsubsection{Dealing with Real Annotation Inconsistency}\label{sec:realdata}
Besides studying the artificial label noise mentioned in previous experiments, it is also important to evaluate the model's performance using a dataset with real annotations from multiple annotators. We used the MSSEG-2016 dataset as an example, which was annotated by seven annotators in parallel. We identified three annotators from the three centers whose labels diverge the most to represent the label noise (\ie, annotator 4 of center 1, annotator 6 of center 7, and annotator 7 of center 8) and conducted the same experiments described in Section~\ref{sec:performance}. The results are reported in Table~\ref{tab:realdata}, which demonstrate that both our CELC and DHLC methods outperform other methods in dealing with annotation variability, highlighting the robustness of our methods within the federated learning framework. 

\begin{table*}[]
\centering
\caption{Lesion segmentation performance of CELC when trained on a real noisy dataset.}
\label{tab:realdata}
\begin{tabular}{c|cccc}
\hline
\multirow{2}{*}{Method} 
& \multicolumn{4}{c}{Metrics} \\ \cline{2-5} 
& \multicolumn{1}{c|}{P-Dice $\uparrow$} 
& \multicolumn{1}{c|}{V-Dice $\uparrow$} 
& \multicolumn{1}{c|}{Precision $\uparrow$} & Recall $\uparrow$\\ \hline
Basic U-Net 
& \multicolumn{1}{c|}{0.4364} & \multicolumn{1}{c|}{0.5698} & \multicolumn{1}{c|}{0.4745} & \multicolumn{1}{c}{0.6212} \\ \hline
Label Smoothing~\cite{muller2019does}
& \multicolumn{1}{c|}{0.4965} & \multicolumn{1}{c|}{0.6411} & \multicolumn{1}{c|}{0.5667} & \multicolumn{1}{c}{0.6138} \\ \hline
Label Correction~\cite{wang2021proselflc}
& \multicolumn{1}{c|}{0.4614} & \multicolumn{1}{c|}{0.5920} & \multicolumn{1}{c|}{0.4910} & \multicolumn{1}{c}{\textbf{0.6647}} \\ \hline
ProSelfLC~\cite{wang2021proselflc}
& \multicolumn{1}{c|}{0.4557} & \multicolumn{1}{c|}{0.5925} & \multicolumn{1}{c|}{0.4915} & \multicolumn{1}{c}{0.6513} \\ \hline
PINT~\cite{shi2021distilling}
& \multicolumn{1}{c|}{0.5493} & \multicolumn{1}{c|}{0.6795} & \multicolumn{1}{c|}{0.6090} & \multicolumn{1}{c}{0.5598} 
\\ \specialrule{2pt}{0pt}{0pt}
U-Net+DHLC
& \multicolumn{1}{c|}{0.5768} & \multicolumn{1}{c|}{0.6826} & \multicolumn{1}{c|}{0.6162} & \multicolumn{1}{c}{0.6013} \\ \hline
U-Net+CELC
& \multicolumn{1}{c|}{\textbf{0.5830}} & \multicolumn{1}{c|}{\textbf{0.6984}} & \multicolumn{1}{c|}{\textbf{0.6492}} & \multicolumn{1}{c}{0.5820} \\ \hline
\end{tabular}
\end{table*}

\section{Conclusion}
\label{sec:conclusion}
In this work, we addressed the challenge of multiple sclerosis lesion segmentation with noisy annotations within the federated learning paradigm. 
Our approach aimed to enhance model robustness in the presence of noisy labels, a situation highly prevalent in this application and address privacy concerns that hinder cross-center collaboration. We conducted the first investigation of this problem in the federated learning setting.
To handle the highly imbalanced distribution of lesions and background regions, we proposed a site-level hard label correction method (DHLC). This method treats positive and negative label corrections separately and directly revises the labels of voxels with high-confidence predictions that differ from the annotations. Moreover, to mitigate the risk of overfitting to label noise during site updating in federated learning, we enhanced the label correction by incorporating predictions from the best-performing central model (with an approach referred to as CELC).
Experimental results on the MSSEG-2016 dataset and our in-house dataset (SNAC-MS) demonstrated the effectiveness and robustness of our proposed methods in combating noisy labels during federated learning.

Although the proposed method is effective, this work still has some limitations that need to be addressed in future research. First, the largest dataset used in this work contains 123 samples distributed to four sites, which is smaller than the real application scenarios. It would be useful to evaluate the method on larger-scale datasets with more samples and more sites. Second, while different noise types were considered to simulate the annotation differences, the domain gap among different sites may also arise from variations in data acquisition devices, which is a common occurrence in real applications but not considered in this work. Thus, it is necessary to develop more robust models that are capable of handling both label noise and image contrast differences among sites.

In conclusion, our study contributes to the understanding of multiple sclerosis lesion segmentation with noisy annotations within the federated learning framework. The proposed method demonstrates promising results and opens avenues for further research in federated segmentation tasks without meticulous labels or knowledge of label noise types and levels.

\appendix

\section*{Acknowledgment}
%Avoid expressions such as ``One of us (S.B.A.) would like to thank $\ldots$ .'' Instead, write ``F. A. Author thanks $\ldots$ .'' 
%In most cases, sponsor and financial support acknowledgments are placed in the unnumbered footnote on the first page, not here.
This work was supported in part by the Australian Department of Health under the Australian Medical Research Future Fund MRFAI000085 (GA89125). Besides, this project and related co-authors were supported by the Australian Research Council Grant DP200103223 and CRC-P Smart Material Recovery Facility (SMRF) – Curby Soft Plastics.

\section*{Declaration of Generative AI and AI-assisted technologies in the writing process}
During the preparation of this work the author(s) used ChatGPT in order to polish the English expression. After using this tool/service, the author(s) reviewed and edited the content as needed and take(s) full responsibility for the content of the publication.

% \bibliographystyle{IEEEtran}
% \bibliography{ref}

% % To print the credit authorship contribution details

\bibliographystyle{model1-num-names}
% \bibliographystyle{biostyles/elsarticle-num}
% \bibliographystyle{cas-model2-names}

\bibliography{ref}

% % Biography
% % Here goes the biography details.

% % Here goes the biography details.

\end{document}